\documentclass[conference]{IEEEtran}
\IEEEoverridecommandlockouts
\usepackage{cite}
\usepackage{amsmath,amssymb,amsfonts}
\usepackage{breqn}
\usepackage{algorithmic}
\usepackage{graphicx}
\usepackage{textcomp}
\usepackage{xcolor}
\usepackage[noabbrev, capitalize]{cleveref}
\def\BibTeX{{\rm B\kern-.05em{\sc i\kern-.025em b}\kern-.08em
    T\kern-.1667em\lower.7ex\hbox{E}\kern-.125emX}}
\begin{document}

\title{Optimal activity and battery scheduling algorithm using load and solar generation forecasts}

\author{
\IEEEauthorblockN{Rui Yuan, Nam Trong Dinh, Yogesh Pipada, S. Ali Pourmouasvi}
\IEEEauthorblockA{\textit{School of Electrical and Electronics Engineering} \\
\textit{University of Adelaide}\\
Adelaide, Australia \\
\{r.yuan, trongnam.dinh, yogeshpipada.sunilkumar, a.pourm\}@adelaide.edu.au}
}

\maketitle

\begin{abstract}

In this report, we provide a technical sequence on tackling the solar PV and demand forecast as well as optimal scheduling problem proposed by the IEEE-CIS technical challenge on predict + optimize for activity and battery scheduling. Using the historical data provided by the organizers, a simple pre-processing approach with a rolling window was used to detect and replace invalid data points. Upon filling the missing values, advanced time-series forecasting techniques, namely tree-based methods and refined motif discovery, were employed to predict the baseload consumption on six different buildings together with the power production on their associated solar PV panels. An optimization problem is then formulated to use the predicted values and the wholesale electricity prices to create a timetable for a set of activities, including the scheduling of lecture theatres and battery charging and discharging operation, for one month ahead. The valley-filling optimization was done across all the buildings with the objective of minimizing the total energy cost and achieving net-zero imported power from the grid. 
\end{abstract}

\section{Introduction}
The aim of this competition was to develop an optimal activity and battery scheduling algorithm taking into account the predictions of the baseload, scheduling constraints and cost of electricity for the month of November 2020 based on historical data\cite{1x9c-0161-21}. Python was selected as the programming language to implement the solution, since it offers libraries for forecasting and optimization. Using Python made the interfacing of these two steps easier. At the beginning, the given building load and solar generation data was visualized for obtaining information on missing points and any cyclic or seasonal patterns that could be extracted. It was found that the solar generation data had less missing information whereas the building data was filled with null points and interrupted patterns. Thus, data cleaning methods such as imputation and deletion was used appropriately to make the data usable for prediction study.\par 

It was clear that the building and the solar patterns were completely different; hence two different set of forecasting models were developed. Various research on forecasting techniques show that the ensemble methods outperform individual ones in many cases. Therefore, we used voting regressor from Python sklearn package was used to forecast buildings' load. This regressor model, fits several estimators on the same data-set and then averages them out to get the actual predictions. It was found that tree-based methods like random forest (RF) plus gradient boosting (GB) method gave the highest accuracy for this data-set. Also, to capture the cyclic and seasonal variations of the buildings' load, STL decomposition was incorporated with the above methods to improve prediction accuracy. \par

Solar generation by its seasonal nature, tends to have repeated patterns. As a result, it might be useful to extract the most repeating pattern from the solar time series data and account for variances from the base-line using exogenous variables such as weather data. This repeating pattern is discovered by refined motif (RM) method which is developed by the authors in \cite{yuan2021irmac}. The discovered repeating patterns along with other exogenous variables were fed to a 1D convolutional neural network (1D-CNN) during phase 1 to make predictions. Some studies have shown that over-parameterization of CNNs can lead to better performance, but these are highly time consuming to train \cite{Zerveas2020}. Therefore, Residual Networks (ResNet) were implemented to develop a deeper NN but with faster computation time. The performance of the ResNet model was generally better compared to 1D-CNN.   \par

The second part of the competition was the development of a cost-minimizing scheduling algorithm. To capture the constraints of this scheduling problem, binary variables are necessary, for example at which interval a particular task is active. Thus, mixed integer programming (MIP) was used to model this problem. From the problem description the following challenges were identified:
\begin{enumerate}
    \item Scheduling for one month with 15-minute granularity means vectors of size 2880. Hence, more activities leads to exponentially increasing complexity.
    \item The peak power cost involved a square term making this problem an mixed integer quadratic program (MIQP).
    \item For best economic benefit, it was necessary to schedule all activities within working hours. This also contributes to the peak power term, which is a sizeable chunk of the energy cost.
\end{enumerate} \par
These challenges heavily influenced the tractability of the problem. We also found that the maximum value obtained by scheduling non-recurring activities was approximately \$16000 and this may not be worth the extra cost and computation required to schedule these tasks. Accounting for this, the following two steps were used to simplify the problem: 
\begin{enumerate}
    \item Only recurring activities were modelled in the problem. 
    \item The problem was converted to a mixed integer linear program (MILP) by setting a limit on the peak power term over the month and removing it from the objective.
\end{enumerate} \par

The methodology was hence divided into 4 sub-sections: data pre-processing, building load forecasts, solar generation forecasts and optimal scheduling problem \cite{FRESNO2021}. 
\section{Background}
The proposed optimization solution uses the idea that the objective solution to the linear program (LP) relaxation of a minimizing MIP problem provides a lower bound for the original objective function \cite{Vielma2015}, i.e.,
\begin{equation}
\label{eq:Property}
    z_{MIP} \geq z_{LP}
\end{equation}
where, \(z_{MIP}\) is the objective value for a minimizing MIP problem and \(z_{LP}\) is the objective value of the LP relaxation of the MILP problem. This property was used to approximate one of the stages of the optimization model to make the computation faster.
\section{Methodology}
\subsection{Data pre-processing}
Data pre-processing step mostly revolved around the handling of missing values from all the energy profiles. Multiple abnormal zero generations for the solar systems and a lot of missing points and outliers for some buildings were found during the inspection. These were classified as invalid datapoints acting as noise while making time-series predictions. To further process these invalid datapoints, three actions were taken:

\begin{enumerate}
    \item Long periods ($>$96 consecutive points) of missing/invalid data were removed from the dataset completely.
    \item Short periods of missing data were approximated using a 96-point moving average window when most data is valid in the window.
    \item When there were more than half the window size of invalid datapoints, these were approximated using the annual average at that time instant. \\
\end{enumerate}

These methods were not used for buildings 4 and 5 because of their peculiar patterns. In building 4, we noticed that the data was in discrete steps of 1, 2, 3 and 4 kW and their appearance seemed to be irrespective of the exogenous variables. Therefore, a flat prediction of 1 kW was applied for the predicting period upon checking the performance of the other discrete values. In building 5, a certain triangular wave pattern with seasonal peaks were noticed. Thus, missing points in September and October 2020 were filled using their corresponding time instant values in 2019. Additionally, in Solar 3, we noticed an increase in the overall generation. This was also accounted for by using Solar 2, which had a similar generation pattern after the increase. Precisely, historical data before May 20\textsuperscript{th}, 2020 of Solar 2 profile were used to replace for those of Solar 3. After filling the missing values, the data pre-processing was completed by adding exogenous variables such as weather information, occupancy, seasonal variations using half sinusoidal waves for solar patterns, climate info (summer or winter) for building load data and day type (weekday or weekend). \par

\subsection{Building load forecasting model}

\begin{figure}[!ht]
\centerline{\includegraphics[clip ,width=0.5\textwidth]{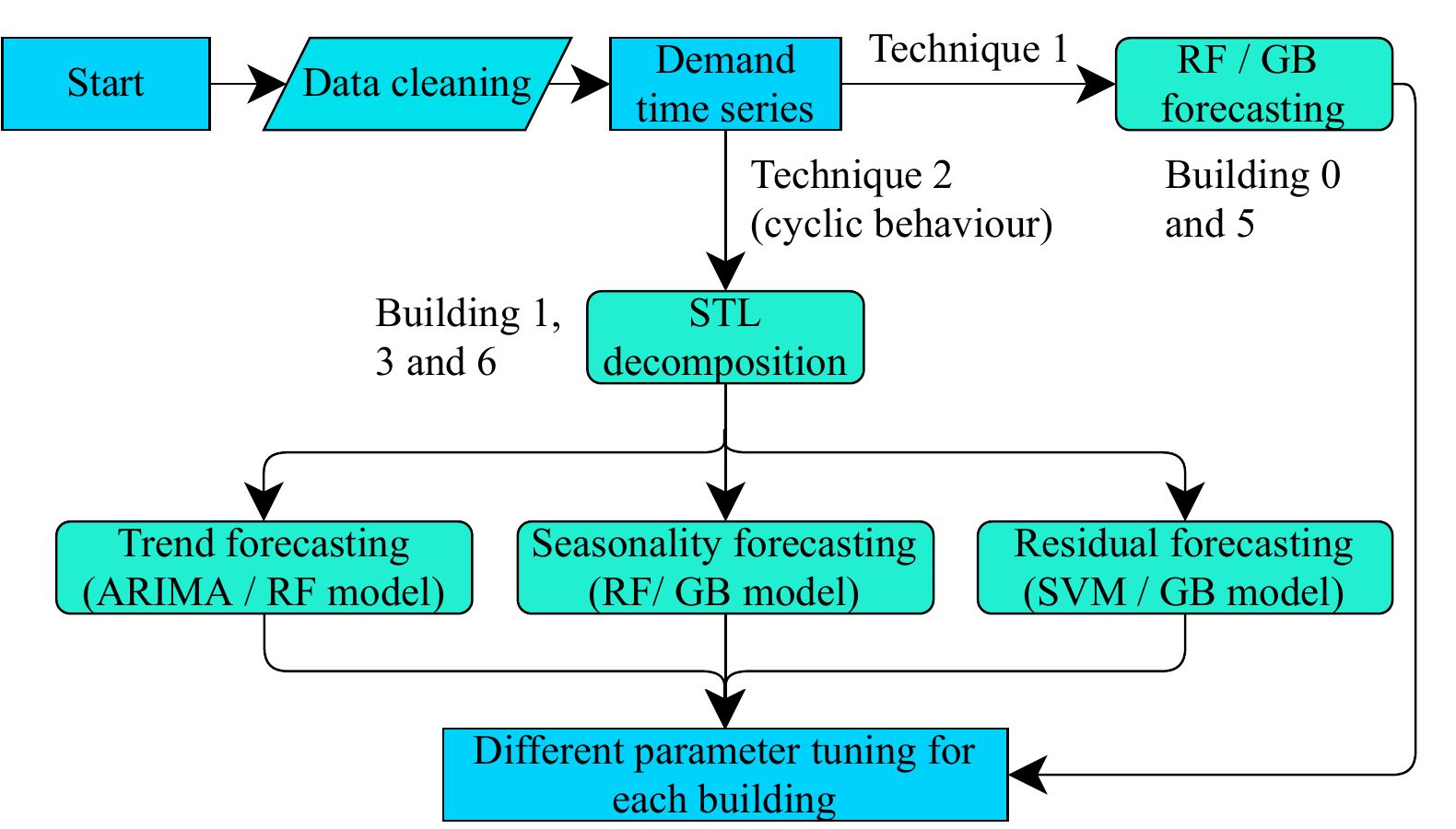}}
\caption{Block diagram for proposed prediction model on building demand}
\label{fig: demand-block-diagram}
\end{figure}

\cref{fig: demand-block-diagram} shows the block diagram of the proposed building forecasting method. Due to the demonstration of cyclic and seasonal behavior from the demand profiles, the concept of seasonal decomposition was applied to enhance the prediction improvement from the tree-based methods. However, for building 0 and 5, the tree-based techniques were used directly on the original profiles as they had a clear and repetitive pattern. These methods include Random Forest (RF) and Gradient Boost (GB). An extra decomposition step was necessary for the other buildings to capture the seasonality, trends, and stochasticity of their time series. Specifically, STL decomposition technique \cite{robert1990stl} was employed to decompose the original time series with 7 days period (96 x 7 intervals). Different regression techniques, including Support Vector Machine (SVM), RF, GB and Autoregressive Integrated Moving Average (ARIMA) were then utilized for training on these components depending on their performances on the validation data in October. Moreover, from the observation of the real data, it could be seen that the demand profiles on Oct 23\textsuperscript{rd} for building 1, 3 and 6 were significantly lower than their expected mean. Since this was the only holiday in the training set, it was not worth having a separate label for this date. Instead, the profiles were replaced with their forecasted values to be later treated as training data for the prediction of November. There was one more holiday in the forecasting period, which appeared on Nov 3\textsuperscript{rd}. Since there was no holiday label, this was accounted for by halving the prediction at mid-day. Similar to the building data, the solar profiles had repeated patterns, which are extracted prior to prediction, as explain in the following section. \par 
\subsection{Solar generation forecasting model}

\begin{figure}[!ht]
\centerline{\includegraphics[clip, trim=0.5cm 16.5cm 5.5cm 0.5cm,width=0.44\textwidth]{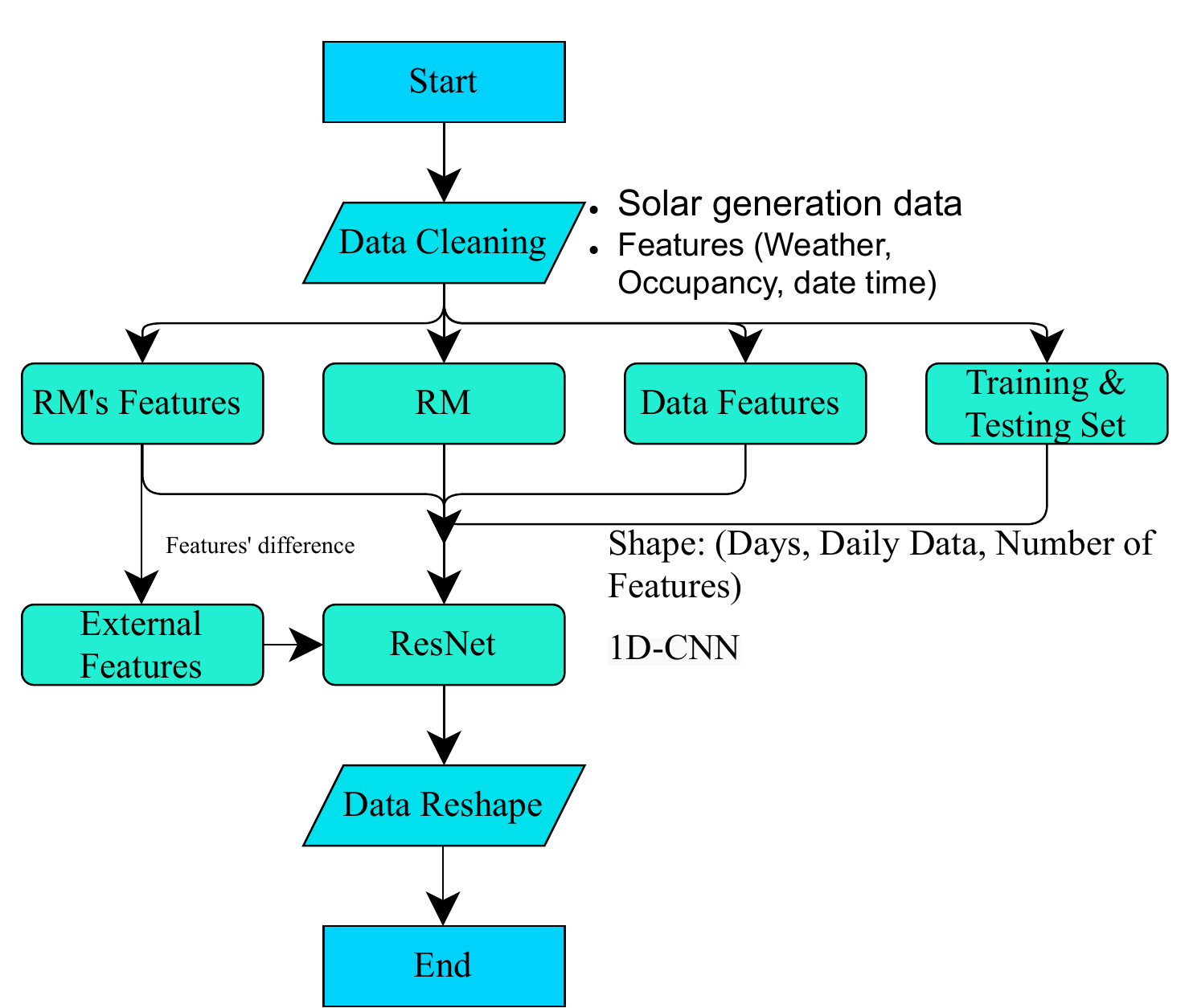}}
\caption{Block diagram for proposed prediction model on solar generation}
\label{fig: solar-block-diagram}
\end{figure}

\cref{fig: solar-block-diagram} shows the block diagram of the proposed solar prediction method, where the pre-processed data provides the daily basis input features including surface radiation, cloud coverage, temperature, and a monthly cycle for presenting the seasonality. The RM is extracted from the generation data by calculating the pairwise similarities of all sub-patterns, and the input features are further processed to get the difference to the RM external features before reshaping it into a daily cycle. Hence, the input features are the representative solar generation and the relative features to the representative model. The output is the solar generation data in a daily cycle with the total prediction length of one month. ResNet is used as the prediction method to facilitate deeper understanding on reshaping the solar generation based on the external features' changes. The hyper-parameters and shortcut connections of the ResNet are chosen based on classic ResNet-18 to ResNet-101 structure, which are originally designed for 2D-CNNs. Adam and MSE are used as the optimizer and error metric, respectively, for the model training. \par

\subsection{Optimal scheduling problem}
\begin{figure}[!ht]
\centerline{\includegraphics[clip, width=0.22\textwidth]{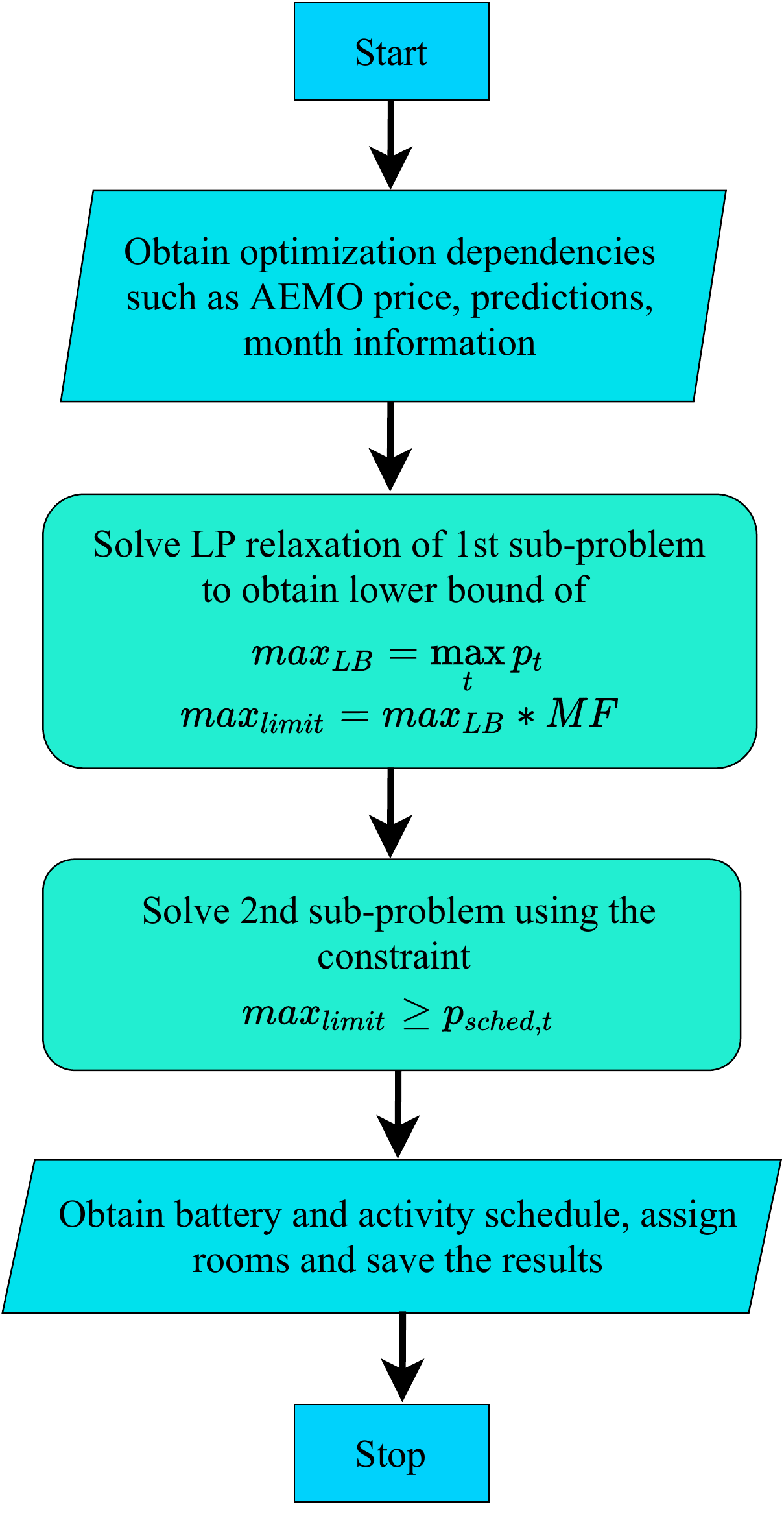}}
\caption{Block diagram for the proposed optimization model}
\label{fig: optimization-block-diagram}
\end{figure}
The scheduling problem was divided into 2 sub-problems; the first sub-problem helps to figure out the limit on peak power and the second one was developed to schedule the activities. The constraints related to the activities are modeled using a day list, which specifies day of the week at a given time index, a non-start list, which specifies non-office hours time indices, two binary decision variable lists to map the start time index and active time indices for each task. The battery constraints are modeled using two binary decision variables; the charge status and discharge status of a battery at some time index. The "battery decision" variable is also calculated in the optimization using a linear combination of these two binary variables. The baseload is calculated from the forecasts obtained from the above methods and used to calculate the power scheduled at a given instant (\(p_{sched,t}\)). Sub-problem 1 involves all the above constraints while minimizing the peak power consumed throughout the month. Whereas, sub-problem 2 aims to minimize the energy costs by limiting \(p_{sched,t}\) using the objective value from sub-problem 1 and scheduling batteries and activities accordingly. The idea here is to get a sub-optimal solution in a reasonably short time. \par
\cref{fig: optimization-block-diagram} shows the sequence of the scheduling problem. First all the dependencies such as AEMO electricity price, load and solar generation forecasts (combined baseload), month information (day list, non start list) and the instance data are obtained. Then the LP relaxation of sub-problem 1 is solved, which gives us a lower bound on the peak power (\(max_{LB}\)) for the given instance. The reason for doing so is to reduce computation time, since the objective value is only of interest and not the schedule itself from this problem. By using \cref{eq:Property}, it is understood that this is the lower bound for the limit; hence multiplied by a sufficiently large multiplication factor (MF) to make sub-problem 2 feasible. This multiplication factor inherently trades off speed for lower objective because the tighter the limit on \(p_{sched,t}\), the slower the optimization will be. \par
The optimization problem was run on a Windows machine using Python 3.8.8 with an Intel i7 8-core processor and 8 GB RAM. Gurobi solver was used to run the optimization. The advantage of using this solver is that it has built-in functions to relax MIP to LP and a well documented Python API. By using trial and error, it was found that a multiplication factor of \(1.10\) and \(1.15\) led to average solution times of \(875s\) and \(1800s\) per instance for small and large instances, respectively. 
The activity and battery schedules were then passed through a room assignment algorithm and finally the solution was saved to the expected format.\par
\begin{figure}[!ht]
\centerline{\includegraphics[width=0.55\textwidth]{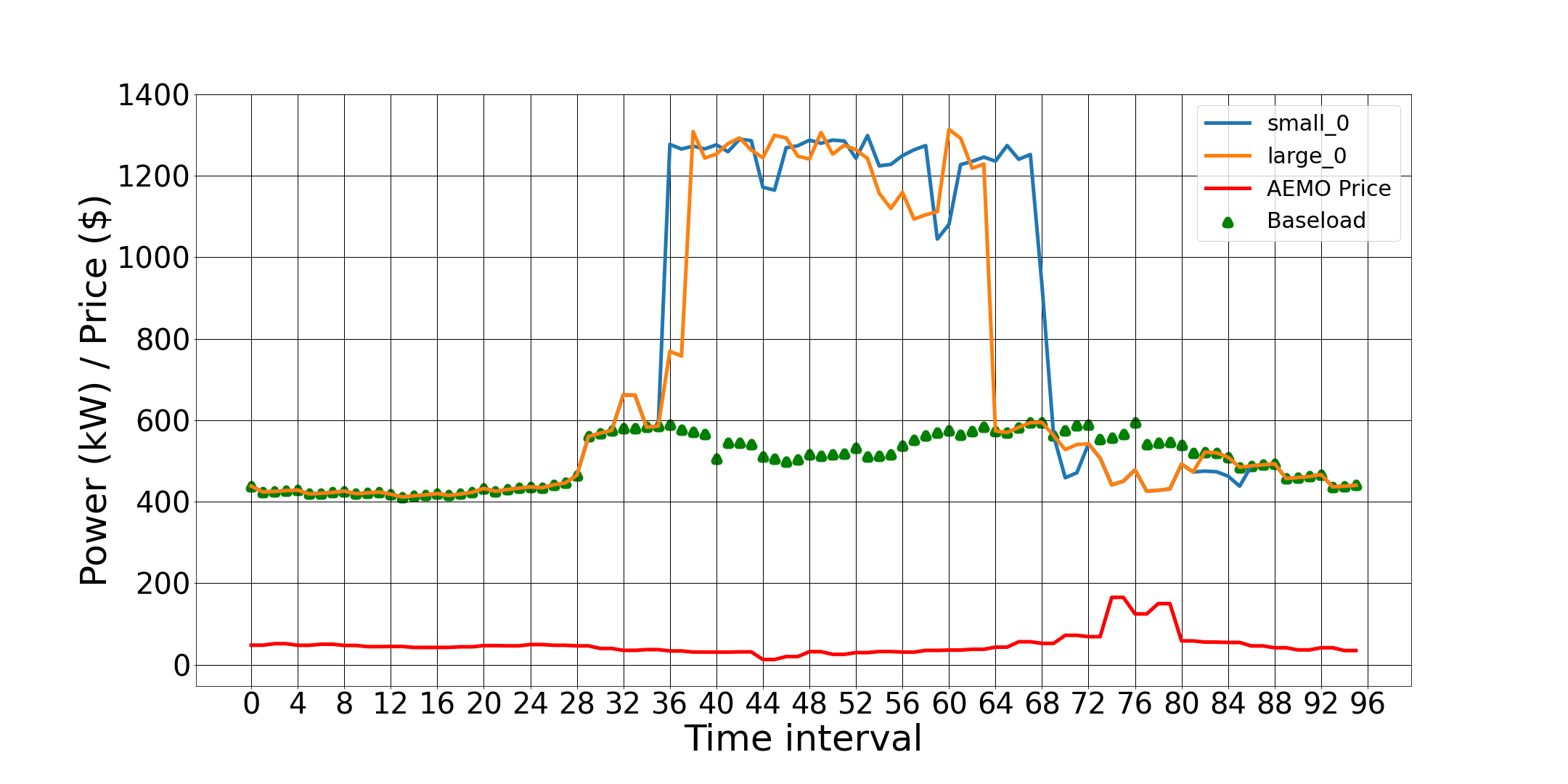}}
\caption{Example power profile from proposed scheduling algorithm}
\label{fig: Results}
\end{figure}

 
\cref{fig: Results} shows an example day, power scheduled for a small and a large instance, along with the baseload (forecasts) and AEMO price. It can be seen that the power profile has a flat peak because of the constraints. It can also be seen that at intervals 72 to 80 when the AEMO price spikes, the scheduled power is less than the baseload. This indicates the batteries are discharged at those times to lower the cost of energy. \par
This method also leads to robust solutions with respect to forecast accuracy, since the problem focuses on minimizing the peak power consumed. Thus, the schedule developed looks to pack all the events as tightly as possible, and even for relatively bad forecasts leads to good results. If the forecasts capture the general trend of the baseload, this method will yield good results because it fills valleys to reach a flat trend. This can also be verified based on phase 2 results submitted, where the forecasts with MASE 1.00 and 1.87 produced schedules with very close objective values.

\section{Experiments}

At the beginning, since most of the training profiles displayed a cyclic behavior with a few noises and outliers, tree-based methods like RF and GB were chosen to provide a rough guideline on the general performance due to their resistance to outliers. Indeed, we achieved an MASE of 0.78 in our very first complete submission in phase 1. From this base-case, various modifications had been made to the inputs, training periods and forecasting techniques on all the profiles as explained above, which then gave us an MASE of 0.67 at the end of phase 1. In phase 2, since there was no feedback from the competition, we decided to build a simplified evaluation function using 30 days' shifted data to justify our improvements. This simplified evaluation function provided a slightly different value to MASE adopted by the competition but was able to reflect the same error changes for predicting values. Under our evaluation function, our models showed improved accuracy from 0.82 to 0.70 for the phase 1 calibration (from simple RF to the new adjustment in phase 2). \par


For solar prediction, we firstly used a two-layer 1D-CNNs as the regression model in phase 1, which provided us with a MASE outperforming the RF and other candidates. Since the evaluation method is MASE, we firstly used MAE and MSE as the NN error metrics instead of the more popular MAPE. The reason is twofold: First, the MAE and MSE are closer to MASE. Second, the MAPE is arguably problematic to outlying errors. With more experiments, we found MSE can lead to a better result than MAE so the former was picked to be the final error metric. Some recent research shows that CNN-based structures can benefit from increasing adjustable parameters, called over-parameterization, whereas one most recent research shows the over-fitting can end up with a more accurate model with sufficient training process. This phenomenon is known as grokking \cite{Power2021} and also discussed as deep double decent in other ResNet related research. To increase the available parameters, we replaced the 1D-CNN with a ResNet for better performance. During the testing to predict October's consumption, this new structure was observed to have a slightly better accuracy for some solar systems. However, due to the lack of November's data and the bad prediction accuracy for our prediction solution in phase 2, using ResNet as the regression model became an unverified improvement. For the sake of training speed, only ResNet-18 and ResNet-34 are tested with early stopping set as 1000 patience, which might not be enough for reaching the Grokking phenomenon. The proposed solar prediction method showed a better accuracy than RF or other state-of-the-art methods in phase 1 and phase 2. However, we did not apply this on buildings consumption prediction because of two reasons: First, the RM based prediction model has a daily cycle and the solar generation is relatively independent between days. Second, learning the building features requires deep NN, which is computationally expensive and requires significant time for thorough testings. \par
\subsection{Conclusion}
The prediction and optimization model was developed and implemented in Python. It was found that the scheduling algorithm was robust with respect to forecasting error. The scheduling of batteries and activities were also visualized to demonstrate the performance of the algorithm.

\nocite{*}
\bibliographystyle{IEEEtran}
\bibliography{reference.bib}

\end{document}